\documentclass[pmlr,twocolumn,10pt]{jmlr}
\pdfoutput=1

\usepackage{longtable}
\usepackage{booktabs}
\usepackage{xcolor}
\usepackage{siunitx}
\usepackage{graphicx}
\usepackage{caption}
\usepackage{algorithm2e}
\usepackage{multirow}
\usepackage{svg}
\usepackage{float}
\usepackage{slashbox}
\usepackage[switch]{lineno}

\usepackage{booktabs}
\usepackage{siunitx} 

\theorembodyfont{\upshape}
\theoremheaderfont{\scshape}
\theorempostheader{:}
\theoremsep{\newline}

\SetKwComment{Comment}{$\triangleright$\ }{}

\title{Multimodal Sleep Apnea Detection with Missing or Noisy Modalities}

 \author{\Name{Hamed Fayyaz} \Email{fayyaz@udel.edu}\\
 \addr University of Delaware
 \AND
 \Name{Abigail Strang} \Email{Abigail.Strang@nemours.org}\\
 \addr Nemours Children's Health
 \AND
 \Name{Niharika S. D’Souza} \Email{niharika.dsouza@ibm.com}\\
 \addr IBM Research Almaden
 \AND
 \Name{Rahmatollah Beheshti} \Email{rbi@udel.edu}\\
 \addr University of Delaware
 }

\begin{document}

\maketitle

\begin{abstract}
Polysomnography (PSG) is a type of sleep study that records multimodal physiological signals and is widely used for purposes such as sleep staging and respiratory event detection. Conventional machine learning methods assume that each sleep study is associated with a fixed set of observed modalities and that all modalities are available for each sample. However, noisy and missing modalities are a common issue in real-world clinical settings. In this study, we propose a comprehensive pipeline aiming to compensate for the missing or noisy modalities when performing sleep apnea detection. Unlike other existing studies, our proposed model works with any combination of available modalities. Our experiments show that the proposed model outperforms other state-of-the-art approaches in sleep apnea detection using various subsets of available data and different levels of noise, and maintains its high performance (AUROC$>$0.9) even in the presence of high levels of noise or missingness. This is especially relevant in settings where the level of noise and missingness is high (such as pediatric or outside-of-clinic scenarios).
\end{abstract}

\paragraph*{Data and Code Availability}
Our code is publicly available at \url{https://github.com/healthylaife/apnea-missing-modality}. The datasets used in this study are available at \url{https://sleepdata.org}.

\paragraph*{Institutional Review Board (IRB)}
This research did not require IRB approval, as it only needed secondary use of deidentified data, per the data owner's (NIH/NHLBI) policy.

\section{Introduction}
Sleep is essential in maintaining and promoting overall health and well-being \citep{luyster2012sleep}. Insufficient or poor-quality sleep has been linked to a wide range of health problems, including cardiovascular disease \citep{kasasbeh2006inflammatory}, obesity \citep{taheri2006link}, diabetes \citep{knutson2006role}, and mental health disorders \citep{zimmerman2006diagnosing,schwartz2005symptoms}.

Conditions that affect sleep quality, timing, or duration and impact a person's ability to function properly while awake are referred to as sleep disorders. There are various types of sleep disorders, including insomnia, circadian rhythm sleep disorders, sleep-disordered breathing, hypersomnia, parasomnias, and restless legs syndrome \citep{pavlova2019sleep}.

In particular, sleep apnea and hypopnea syndrome (SAHS) are breathing disorders that are characterized by recurring pauses in breathing during sleep, which usually cause fragmentation of sleep and can lead to oxygen deprivation and disrupt normal physiological processes \citep{vaquerizo2020automatic}. 
The main types of SAHS are obstructive sleep apnea (the most common type), central sleep apnea, complex (mixed) sleep apnea, and hypopnea. It is estimated that 26\% of people between 30 and 70 years \citep{schwartz2018effects} and 1\% to 5\% of children suffer from SAHS, with the highest prevalence in age 2 to 8  \citep{kheirandish2012sleep,bixler2009sleep, marcus2012diagnosis}. 

Polysomnography (PSG) is the gold standard for diagnosing sleep-related breathing disorders. It refers to the process used to collect biological signals and parameters during sleep, which is generally performed in clinical lab settings and during the night \citep{rundo2019polysomnography}. A PSG generally includes: 1) brain electrical activity (electroencephalogram or EEG), 2) eye movements during sleep (electrooculogram or EOG), 3) cardiac rate and rhythm (electrocardiogram or ECG), 4) blood oxygen saturation (pulse oximetry or SpO\textsubscript{2}), 5) measurement of exhaled air to indirectly measure blood CO\textsubscript{2} (end-tidal carbon dioxide or ETCO\textsubscript{2}), 6) respiratory effort in thorax and abdomen (respiratory inductance plethysmography or RIP), 7) and nasal and oral airflow. PSG is generally considered effective; however, it presents many challenges, including complexity, cost, intrusiveness, and the need for intensive involvement of clinical providers \citep{spielmanns2019measuring}. The multimodal nature of PSG data provides a diverse and holistic view of the subjects \citep{muhammad2021comprehensive, kline2022multimodal, 10.1007/978-981-19-9865-2_10}. 

While a fairly large family of studies aiming to detect SAHS from PSG is present, existing methods have two major limitations. First, conventional methods are built based on the assumption that all modalities are available for all subjects. However, 
in real-world scenarios, certain modalities may be partially or entirely missing due to technical issues or limitations during data acquisition. Additionally, the PSG signals can be corrupted by noise from various sources, including electrode artifacts, electrical interference, or patient movement. Detecting apnea becomes significantly more challenging when dealing with missing or noisy modalities (especially the primary signals such as SpO\textsubscript{2}).
The second limitation is the applicability of existing methods to a predetermined subset of PSG signals. This issue makes it unclear to what degree a method developed for a certain set of modalities would work (if at all) on applications with different sets of available modalities. To address these limitations, we propose a flexible machine learning-based pipeline for apnea detection with any combination of available signal modalities. Specifically, the contributions of our study are:

\begin{itemize}
    \item We extensively investigate the effects of missing or partially available modalities in apnea detection.
    \item We present a novel universal pipeline to predict apnea with any type, length, or quality of available modality. 
    \item Through extensive experiments, we show that our method is robust to noisy and missing modalities and outperforms prior methods in various apnea detection scenarios. 
\end{itemize}

\section{Related Work} \label{sec:background}

Due to the high relevance of detecting apnea events using machine learning methods, many attempts have been made to automate apnea detection using such methods. A large body of prior work focuses on uni-modal methods. Considering the modalities utilized in prior studies, we highlight four categories in the following (a non-exhaustive list). (1) Most existing work uses ECG for apnea detection. while ECG is not the most relevant signal in clinical settings, the popularity of the methods for ECG is partly due to the availability of public ECG datasets \citep{penzel2000apnea}. These methods generally use `band-pass' filters to reduce the noise sourced from the baseline wander, muscle artifacts, power line interference, and other sources \citep{urtnasan2018automated, bahrami2022sleep}.  Since ECG signals contain complex patterns, many studies have used extensive preprocessing steps, including automatic feature extraction approaches through deep neural networks, in their pipeline for extracting features \citep{shen2021multiscale, 
chang2020sleep, chen2022toward, zarei2022detection}. These methods are extensively reviewed by \cite{salari2022detection}. (2) Many studies have used SpO\textsubscript{2}, which is the most clinically relevant signal for apnea detection (in adults). Similar to the previous category, various types of manual or automatic feature extraction methods have been used in this category \citep{alvarez2012feature, morillo2013probabilistic, uccar2017automatic, john2021somnnet}. (3) The combination of these two signals (ECG and SpO\textsubscript{2}) have been used, especially to handle the signals' imperfection and defects, such as missing data or noise \citep{tuncer2019deep, ravelo2015oxygen, pathinarupothi2017single, xie2012real}. (4) Finally, the fourth major category is related to a group of work that uses EEG signals to detect apneic events \citep{vimala2019intelligent, zhao2021classification, almuhammadi2015efficient}. 
As discussed earlier, a common limitation of existing studies is unclear generalizability to other PSG signals.  
Oftentimes, these models cannot utilize an additional modality not seen in the training phase and handle partially available or noisy modalities.

Multimodal learning approaches aim to use information from different sources to better understand an underlying phenomenon and improve the performance of downstream tasks. Data fusion is an integral part of multimodal learning, which is the process of integrating multiple data sources to build a representation that is more pertinent than any individual source \citep{stahlschmidt2022multimodal}. 
Depending on the fusion stage, existing techniques can be categorized into i) early, ii) intermediate, and iii) late fusion, where the data from modalities are respectively combined at i) the input \citep{lim2019radar, neves2023end}, ii) through a middle representation (like a hidden layer) between the individual modalities and the final model \citep{hassan2022early}, and iii) using separate models on each modality which are combined later \citep{wang2019multi, zhang2019late}. When some modalities are missing or noisy, data fusion enables combining available information from different sources to create a more comprehensive and reliable representation, enhancing the robustness of the models \citep{gaw2022multimodal}.

Multimodal information can be especially effective for apnea detection, as exploiting multimodal data not only can outperform using only uni-modal data but can also improve the robustness of the system when one or more modalities are missing. Despite the importance of multimodal learning, there is limited work \citep{ye2023diagnosis, fayyaz2023bri, pmlr-v116-van-steenkiste20a} on multimodal data fusion for apnea detection. We are not aware of any previous study that directly aims at handling missing and noisy data in the context of sleep apnea detection. 

When it comes to working with multi-modal data, the choice of a data fusion mechanism presents an additional dimension of the problem. In machine learning pipelines, fusing multimodal data in any of the discussed stages has advantages and disadvantages. Among the existing fusion techniques, gated fusion \citep{arevalo2017gated} is a popular method, with applications in different domains such as computer vision \citep{hosseinpour2022cmgfnet, zhang2018gated, ren2018gated, feng2020cpfnet}, speech recognition \citep{fan2020gated}, and natural language processing \citep{du2022gated}. By employing gating mechanisms, gated fusion models selectively weigh and fuse information and capture complementary aspects. In our study, we propose a modified version of the gated fusion mechanism that considers the abnormality of the present modalities (i.e., the extent to which current signal patterns deviate from the others)  to identify the optimized way of ``fusing'' the present modalities.

\section{Problem setup}
Consider a sleep dataset consisting of ${I}$ sleep studies (e.g.,  ${I}$ individual nights in the clinic) shown by $\{S_{i}\}_{i=1}^{I}$. Each study can be divided into equal-length non-overlapping epochs (windows), considered as a sample $X$, and shown by

\begin{equation}
\label{eq:1}
  S_{i} = \{X^{i, j}\}_{j=1}^{J_{i}},
\end{equation}
\noindent
where, $J_{i}$ is the number of epochs in  $i$-th study.

Each sample can consist of different PSG signals, considered as distinct modalities:
\begin{equation}
  X^{i,j} = (X^{i,j}_{1}, X^{i,j}_{2}, ..., X^{i,j}_{M}),
\end{equation}

\noindent
where $M$, is the total number of modalities. 

Furthermore, each sample has a label $y$, showing if an apnea event occurs during the sample time window.  Thus, $X^{i,j}_{m}$ is a signal from the modality $m$ that belongs to the epoch $j$ of the study $i$, and is a multivariate time series consisting of $T$ time points:

\begin{equation}
    X^{i,j}_m = (X^{i,j}_{1,m}, X^{i,j}_{2,m}, ..., X^{i,j}_{T,m}).
\end{equation} 


In this work, we design and train a model $f$ (parameterized with $\theta$) that can handle missing and noisy modalities and estimates the occurrence of apnea $\hat{y}$   in a given epoch:

\begin{equation}
\hat{y}^{i,j} = f_{\theta}(X^{i,j})
\end{equation}


\section{Method}

\begin{figure*}[ht]
    \centering
    \includegraphics[width=1.04\linewidth]{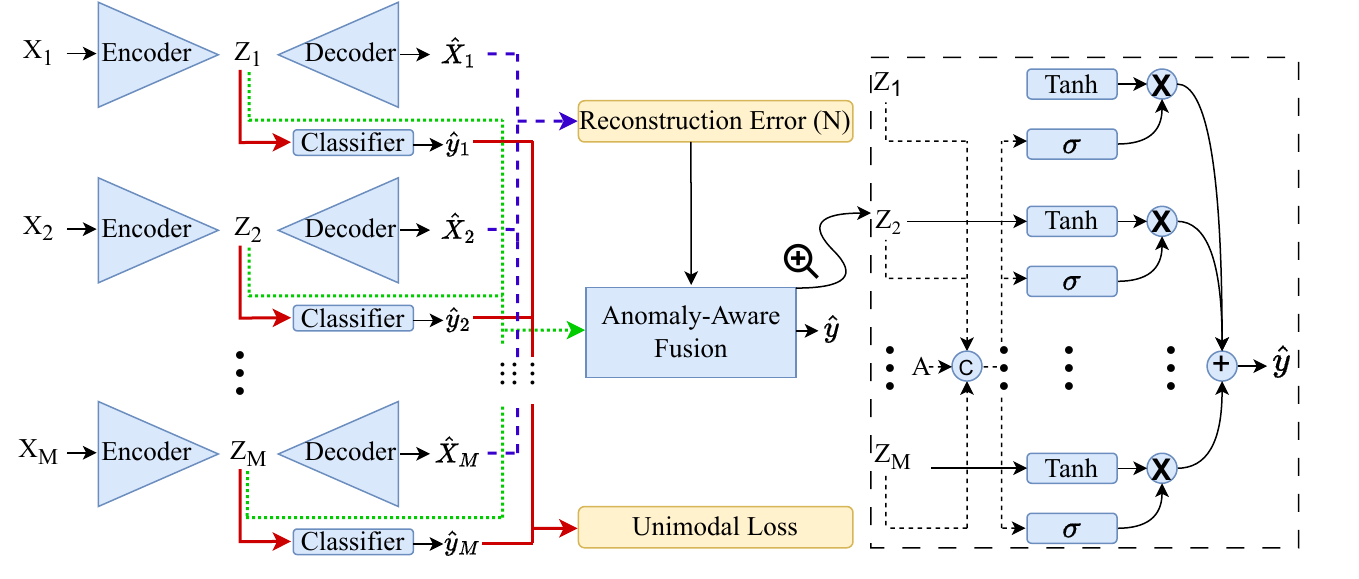}
    \caption{Our proposed pipeline. It consists of four components: (1) Encoder, (2) Decoder, (3) Classifier, and (4) Anomaly-aware Fusion. In the first (unimodal) step of training, an encoder for each modality is used to transform the input into a latent space $Z_{m}$. Besides, a unimodal classifier and decoder are utilized to detect apnea and reconstruct the input, respectively. In the second (multimodal) step of training, the Anomaly-Aware Fusion module classifies the epochs using the unimodal latent representation ($Z$) and the reconstruction error for each input signal. Notations \texttt{C} and \texttt{X} denote the concatenation and multiplication operations respectively.}
    \label{fig:overview}
\end{figure*}

For detecting apnea with missing or noisy modalities, we propose a two-step semi-supervised learning method in companion with a multimodal autoencoder network using a transformer backbone (Figure \ref{fig:overview}). The two steps of our method include unimodal pre-training and multimodal training. 
The rationale for introducing a two-step training process in our design is that the pipeline can use samples that do not have complete modalities to train the model when dealing with incomplete modalities (akin to an imputing mechanism paradigm).

\paragraph{Unimodal Pre-training}
We use unimodal transformer-based autoencoders that reconstruct a unimodal input $X_{m}$, by mapping the input space to a lower dimensional latent space $Z_{m}$. More specifically, an encoder $E_{m}$ maps the signal to a latent representation ($Z_{m} = E_{m}(X_{m})$) and a decoder $D_{m}$ reconstructs the original signal from the latent representation ($\hat{X}_{m} = D_{m}(Z_{m})$).

The encoder and decoder consist of a series of transformer blocks. Besides being used for signal reconstruction, the decoder is also used for anomaly detection by using the reconstruction error as a proxy. 
Our method also includes a unimodal classifier $C_{m}$ to detect the epochs with apneic events using $Z_{m}$ ($\hat{y}_{m} = C_{m}(Z_{m})$).

This way, the encoder learns a mapping to a latent space that preserves the fundamental characteristics of the signals, while extracting the key features for apnea detection. Additional details about the design of the transformers are presented in Appendix \ref{tfb}.

For modality $m$, the reconstruction loss $\mathcal{L}_{R}$ can be formulated as the mean squared error ({\tt MSE}) of the original signal and its reconstructed version:

\begin{equation}
\label{rec-err}
    \mathcal{L}^{m}_{R} =\frac{\sum_{i=1}^{I}\sum_{j=1}^{J_i}\sum_{t=1}^{T}(X^{i,j}_{t,m} - \hat{X}^{i,j}_{t,m})^2}{N \times T},
\end{equation}

\noindent
where, 

\begin{equation}
   N =  \sum_{i=1}^{I}J_i, \text{ and}
\end{equation}

\begin{equation}
    \hat{X}^{i,j}_{t,m} = D_{m}(E_{m}(X^{i,j}_{t,m})).
\end{equation}

In addition, the classification loss $\mathcal{L}_{C}$ (related to the apnea detection task) is defined using binary cross entropy ({\tt BCE}) for modality $m$ as:

\begin{equation}\label{eq:loss}
    \mathcal{L}^{m}_{C} = \frac{1}{N}\sum_{i=1}^{I}\sum_{j=1}^{J_i}BCE(y_{m}^{i,j},\hat{y}_{m}^{i,j}),
\end{equation}

\noindent
where, 

\begin{equation}
    \hat{y}^{i,j} = C_{m}(E_{m}(X^{i,j}_{m})).
\end{equation}

The model (in this first step) is trained by jointly optimizing the  two aforementioned loss functions over all modalities:
\begin{equation}
    \mathcal{L}_{Unimodal} = \sum_{m=1}^{M}\alpha_{m} \mathcal{L}^{m}_{R} + \beta_{m}  \mathcal{L}^{m}_{C}
\end{equation}
where, the hyperparameters ($\alpha_{m}$ and $\beta_{m}$) are used to tune the contribution of each loss.

\paragraph{Multimodal Training}

In the second step of our method, we use a modified version of gated fusion \citep{arevalo2017gated} to detect apneic events. We refer to this module as Anomaly Aware Fusion ({\tt AAF}). It receives the latent representation $Z$ and a distance measure $A$ for each signal. The distance measure $A$ aims to capture the abnormality of the signals (the degree to which the signal deviates from the original), similar to \cite{anovit}. Here, $A$ is defined as:
\begin{equation}
    A = [a_{1}, a_{2}, ..., a_{M}],
\end{equation}

\noindent where, 
\begin{equation}
\label{rec-loss}
    a_{m}=[|X_{1,m}-\hat{X}_{1,m}|,..., |X_{T,m}-\hat{X}_{T,m}|],
\end{equation}
\noindent
while the anomaly score $a_{m}$ captures the absolute distance between the two time-series  $X_{m}$ and $\hat{X}_{m}$.  While closely related, the reconstruction loss in Eq. \ref{rec-loss} yields a single value; however, the distance measure in Eq. \ref{rec-err} is a vector that shows the degree of abnormality in each time step.

Specifically, {\tt AAF} can be shown as a function receiving $Z$ and $A$ to predict the final label:

\begin{equation}
    \label{eq:naf}
    \hat{y} = AAF(Z, A).
\end{equation}

As the encoder-decoder model learns the distribution of (mostly normal) signals, the distance measure (captured by the absolute differences) increases when this model receives an abnormal signal, as the signals containing the abnormal patterns are not reconstructed well.

Furthermore, $Z$ in {\tt AAF} (Eq. \ref{eq:naf}) refers to:
\begin{equation}
    Z = [Z_{1}, Z_{2}, ..., Z_{M}],
\end{equation}
\noindent
where, $Z_m$ corresponds to a feature vector in the latent representation space associated with the modality $m$. 

The {\tt AAF} module (the right box in Figure \ref{fig:overview}) takes these steps. First, each feature vector $Z_{m}$ is fed to a fully connected layer with a $Tanh$ activation function, which is intended to encode an internal representation feature $h_m$ based on the particular modality: 

\begin{equation}
    h_{m} = Tanh(W_{m} Z_{m}).
\end{equation}

\noindent
where, $W_m$s are learnable weights related to the modality $m$. Another fully connected layer is used for each $Z_m$, which controls the contribution of the feature calculated from $Z_m$ to the overall output of  {\tt AAF}. The output of  this layer is another latent representation named $Z^{'}$:

\begin{equation}
    Z^{'}_{m} = \sigma(W_{Z_{m}} [Z_{1}, Z_{2}, ..., Z_{M}, A])
\end{equation}

\noindent
where, $W_{Z_{m}}$s are learnable weights related to modality $m$.

When a sample is fed to the gated fusion network, a gate layer associated with the modality $m$ receives the feature vectors of all modalities and uses them to decide whether the modality $m$ may contribute or not to the internal encoding of the particular input sample:

\begin{equation}
    h = Z^{'}_{1}h_{1} + Z^{'}_{2}h_{2} + ... + Z^{'}_{M}h_{M}.
\end{equation}

Finally, a fully connected  layer and a sigmoid function $\sigma$ transform $h$ into a label:

\begin{equation}
    \hat{y} = \sigma((W_{c}h) + b_{c}),
\end{equation}
where, $W_c$ and $b_c$ are learnable weights and biases.  We use the {\tt BCE} loss (as in the previous step) for the classification loss. The encoder and decoder weights are frozen during this step.

\section{Experiments}

\paragraph{Datasets}
A large number of public sleep studies are available, including those available through the National Sleep Research Resource
platform \citep{sleepdata}. Most of those, however, are not a good fit for evaluating our method. For our study, we need sleep recordings that (i) have multimodal data (i.e., not unimodal or only a few modalities), (ii) are supervised (i.e., have expert annotated apnea event labels), and (iii) have a large number of samples (the \textit{n} size for the sleep studies and patients), to allow training large deep learning models. We use two of the largest public sleep datasets that match the above criteria. These two are the Nationwide Children's Hospital (NCH) Sleep Data Bank \citep{lee2022large} and Childhood Adenotonsillectomy Trial (CHAT) dataset \citep{marcus2013randomized, redline2011childhood}.

\textbf{NCH} - The first dataset offers a large and free source that includes both PSG signals and linked electronic health records for the patients. The linked data includes demographics and longitudinal clinical data such as encounters, medication, measurements, diagnoses, and procedures.  The dataset was collected between 2017 and 2019 at Nationwide Children's Hospital (NCH), Columbus, Ohio, USA. Sleep studies were annotated in real-time by technicians at the time of the study, and then were staged and scored by a second technician after the study was completed. We used all studies that have all of the available six modalities.

\textbf{CHAT} - We also use recordings from the CHAT study, which is a randomized, single-blind, multicenter trial designed to analyze the efficacy of early removal of adenoids and tonsils (adenotonsillectomy) in children with mild to moderate obstructive apnea. Physiological measures of sleep were assessed at baseline and at seven months with standardized full PSG with central scoring at the Brigham and Women’s Hospital, Boston, MA. In total, 1,447 children had screened PSG and 464 were randomized to treatment. Children underwent standardized PSG testing with scoring at a centralized sleep reading center, cognitive and behavioral testing at baseline, and 7 months after randomization. We use the PSGs collected in the baseline in our work.



\paragraph{Baselines}  We chose three related and state of the art (SOTA) studies with different architectures to compare with the proposed model. These three studies include architectures based on CNNs \citep{chang2020sleep}, CNN+LSTM \citep{zarei2022detection}, and 
transformers \citep{fayyaz2023bri} as described in the following.

\textbf{CNN} - The first study by \cite{chang2020sleep}  uses a network consisting of 10 CNN layers for feature extraction followed by four fully connected layers for classification. They also apply batch normalization and dropout for better generalization and to avoid overfitting. They only utilized ECG for apnea detection. However, to have a fair comparison, we trained this model with the same modalities that we used for training our model.

\textbf{CNN+LSTM} - The second model presented by \cite{zarei2022detection} uses an automatic feature extraction method developed by combining CNN and long short-term memory (LSTM) modules using ECG. A stack of fully connected layers is used at the end to classify the events. Similar to the previous baseline, we trained this model with the same modalities that we used for training our model.

\textbf{Transformers} - The last model proposed by \cite{fayyaz2023bri} uses transformers followed by two layers of a fully connected network as a classifier for detecting apnea using PSG signals.

\paragraph{Implementation details}  Our model utilizes ECG, EEG, EOG, SpO\textsubscript{2}, CO\textsubscript{2}, and respiratory signals to detect apnea. As the signals have different sampling rates, we resampled all signals with a frequency of 128Hz. For model input, we divided each sleep study into 30-second non-overlapping epochs. ECG signal was denoised using a band-pass filter, which allows only a certain range of frequencies to pass through while blocking others, with lower and upper cutoff frequencies of 3Hz and 45Hz. Hamilton R-peak detection method \citep{hamilton1986quantitative} was utilized to extract R-R intervals and the amplitude of R-peaks from the ECG signal.



We run experiments in a 5-fold cross-validation manner and report the mean and standard deviation of performance of the trained models using the F1-score (harmonic mean of precision and recall) and AUROC (area under the receiver operating characteristic curve) in the test set.

\subsection{Research Questions}
We study five major research questions in our experiments to investigate the performance of our model.

\paragraph{Q1: How do the methods perform with complete modalities?}
At first, we carried out an experiment to find the overall performance of our method and the baselines when provided with complete data (i.e., without any added noise or missingness). Table \ref{tab:resultbal} shows the results. Our model has achieved superior performance, measured in terms of F1 score and AUROC, compared to the baselines. We also ran an ablation study to evaluate the performance of our model in the absence of reconstruction loss (in unimodal pretraining) and the distance measures $A$ (in multimodal training).

\begin{table*}[ht]
\centering
\caption{The overall performance when using complete data. The three baselines are CNNs \citep{chang2020sleep}, CNN+LSTM \citep{zarei2022detection}, and 
transformers \citep{fayyaz2023bri}. Mean ($\pm$ STD).}
\label{tab:resultbal}
\begin{tabular}{l|cccc}
\toprule
\multicolumn{1}{c|}{\multirow{2}{*}{Method}} & \multicolumn{2}{c}{CHAT}                      & \multicolumn{2}{c}{NCH}             \\ \cline{2-5} 
\multicolumn{1}{c|}{}                      & F1                  & AUROC              & F1                 & AUROC            \\
\midrule
CNN                  & 77.5 (0.8)          & 86.8 (1.0)          & 77.2 (1.1)          & 86.4 (1.2)        \\
CNN+LSTM         & 81.7 (0.6)          & 89.7 (0.7)          & 81.7 (0.8)          & 89.4 (0.6)          \\
Transformer           & 83.1 (1.0)          & 90.0 (0.8)         & 82.6 (0.5)          & 90.4 (0.4)        \\
\midrule
Ours without reconstruction loss                               & \textbf{85.1 (0.7)}                 & \textbf{92.3 (0.6)}                & \textbf{83.9 (0.6)}               & \textbf{92.3 (0.4)}              \\
Ours                                       & \textbf{86.6 (0.7)}                 & \textbf{93.3 (0.6)}                & \textbf{85.2 (0.7)}               & \textbf{93.4 (0.4)}              \\
\bottomrule
\end{tabular}
\end{table*}

\begin{figure}[h]
  \centering
  \includegraphics[width=1\linewidth]{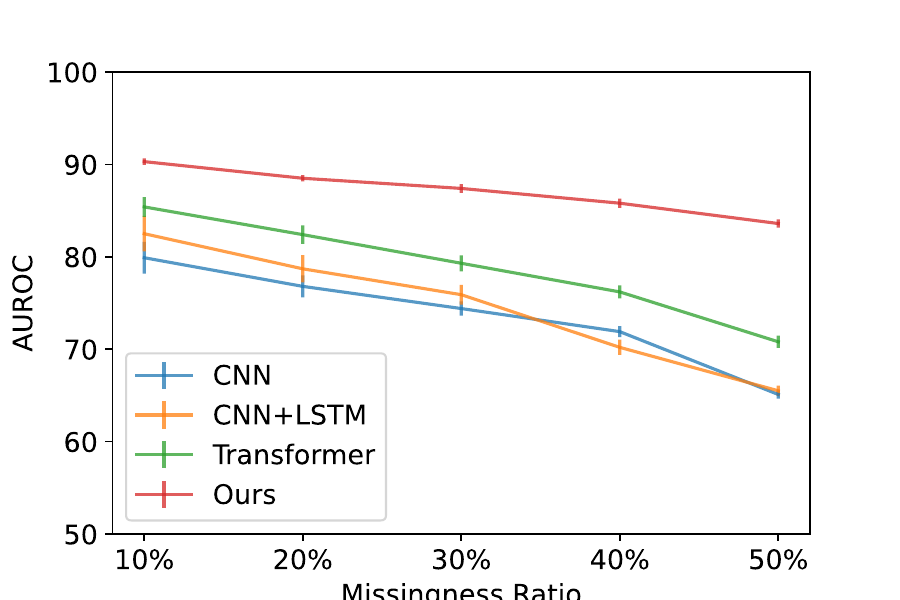}
   \caption{Models Performance with missing modalities on CHAT dataset.}
   \label{fig:missing-chat}
\end{figure}

\paragraph{Q2: How do the methods perform in the presence of missing modalities?}
To simulate the effect of missing modalities, we omitted signals within epochs in a random manner. More specifically, we randomly substituted a predefined percent of signals in epochs with zero. The algorithm for this purpose is presented in Appendix \ref{algs}. Figure \ref{fig:missing} (NCH data) and \ref{fig:missing-chat} (CHAT) show the results.

\begin{figure}[h]
  \centering
  \includegraphics[width=1\linewidth]{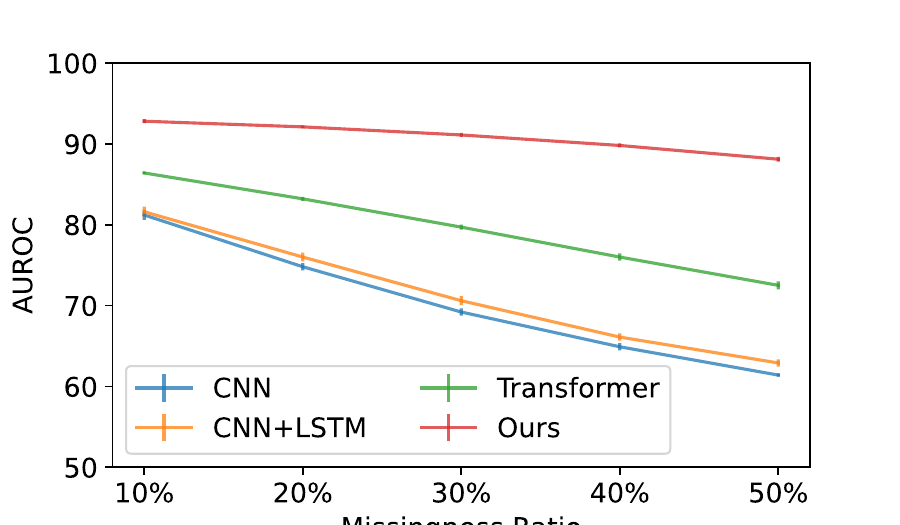}
   \caption{Model performance with missing modalities on the NCH dataset. Error bars show the standard deviation.}
   \label{fig:missing}
\end{figure}

\begin{figure}[h]
    \centering
  \includegraphics[width=1\linewidth]{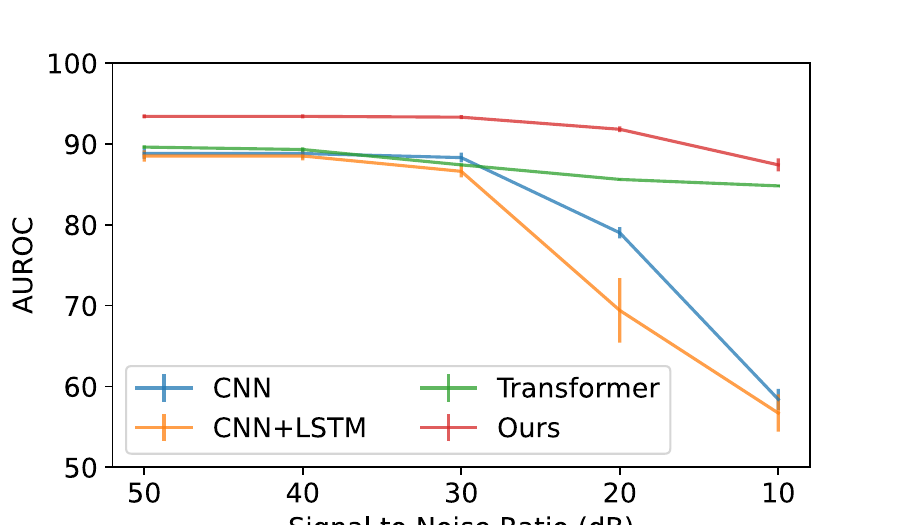}
    \caption{Model performance with noisy modalities on the NCH dataset.}
    \label{fig:noisy}
\end{figure}

\begin{figure}[h]
    \centering
  \includegraphics[width=1\linewidth]{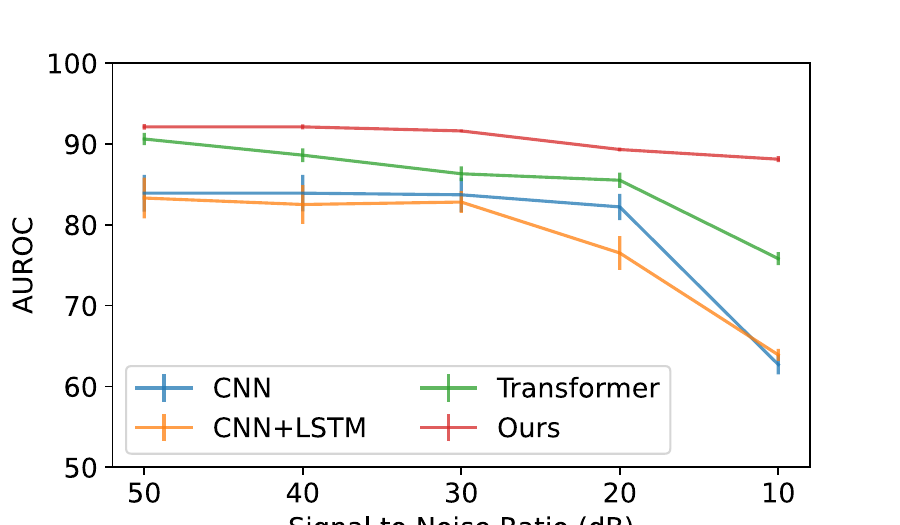}
    \caption{Models performance with noisy modalities on CHAT dataset.}
    \label{fig:noisy-chat}
\end{figure}

\paragraph{Q3: How do the methods perform with noisy modalities?}
We investigated how well our model and other top-performing models handle noisy data. In the absence of a dataset with controlled non-synthetic or measured noise, we add synthetic noise to our target datasets.  We added white Gaussian noise to the signals to simulate unremovable types of noises (we discuss this further in \ref{sec:disc}), using a process explained further in Appendix \ref{algs}. Additive Gaussian noise can mimic the effect of many random processes that can occur in real-life settings. The results are shown in Figures \ref{fig:noisy} (NCH) and \ref{fig:noisy-chat} (CHAT). The signal-to-noise ratio (SNR) is a measure of the strength of a desired signal compared to the background noise. It is often expressed in decibels (dB), and a higher SNR value indicates a cleaner signal.

\paragraph{Q4: How do the methods perform when the input has both noisy and missing signals?}
We also fed our model with both noisy and missing modalities at once, which combines the two studied scenarios in Q3 and Q4. The results on the NCH and CHAT datasets are shown in Table \ref{tab:noise-missing}. We report the results of the baseline models for the Q4's scenario, in the Appendix (Table \ref{tab:noise-missing-sota}).

\begin{table*}[ht]
\centering
\caption{Our method's performance (mean AUROC $\pm$ STD) with the concurrent occurrence of noise and missingness on NCH dataset.} 
\label{tab:noise-missing}
\resizebox{2.05\columnwidth}{!}{
\begin{tabular}{l|llllllllll}
\toprule
              & \multicolumn{5}{c|}{NCH}                                                       & \multicolumn{5}{c}{CHAT}                                       \\ \cline{2-11} 
Missing Ratio & \multicolumn{10}{c}{Signal to noise ratio (dB)}                                                                                                 \\
              & 10        & 20        & 30        & 40        & \multicolumn{1}{l|}{50}        & 10         & 20         & 30         & 40         & 50         \\ \cline{2-11} 
10\%          & 84.9(1.9) & 90.4(0.9) & 92.6(0.5) & 92.8(0.5) & \multicolumn{1}{l|}{92.8(0.5)} & 85.5 (0.7) & 85.5 (0.6) & 90.0 (0.1) & 91.7 (0.4) & 91.8 (0.5) \\
20\%          & 82.9(1.9) & 89.0(1.1) & 91.8(0.6) & 92.0(0.5) & \multicolumn{1}{l|}{92.1(0.5)} & 83.1 (0.7) & 82.6 (1.5) & 87.7 (0.6) & 89.3 (0.1) & 89.9 (0.3) \\
30\%          & 80.9(2.0) & 87.3(1.0) & 90.6(0.6) & 91.1(0.5) & \multicolumn{1}{l|}{91.1(0.5)} & 79.8 (0.8) & 79.7 (0.9) & 85.0 (1.1) & 87.5 (0.1) & 87.3 (0.1) \\
40\%          & 78.6(1.9) & 85.4(1.2) & 89.3(0.6) & 89.9(0.5) & \multicolumn{1}{l|}{89.8(0.5)} & 76.6 (0.5) & 76.8 (1.5) & 81.8 (1.5) & 84.4 (0.8) & 84.6 (0.6) \\
50\%          & 76.9(2.0) & 83.4(1.0) & 87.6(0.6) & 88.0(0.6) & \multicolumn{1}{l|}{88.2(0.4)} & 73.5 (0.1) & 73.6 (1.7) & 78.6 (0.9) & 81.0 (1.1) & 81.1 (0.9)
\end{tabular}
}
\end{table*}

\paragraph{Q5: How do the methods perform when specific modalities are completely missing?}

We also study the methods with one missing modality at a time, to find the stability of our model when modalities are entirely (versus partially in Q2) unavailable. Results are shown in Table \ref{tab:modality-missing}.

\begin{table*}[ht]
\centering
\caption{Performance of our model (mean AUROC $\pm$ STD) when specific modalities are missing on the NCH dataset. }
\label{tab:modality-missing}
\resizebox{2.05\columnwidth}{!}{
\begin{tabular}{l|llll|llll}
\toprule
\multirow{2}{*}{Missing Modality}     & \multicolumn{4}{c|}{NCH}                               & \multicolumn{4}{c}{CHAT}                                 \\ \cline{2-9} 
                                      & CNN        & CNN-LSTM    & Transformer & Ours       & CNN         & CNN-LSTM    & Transformer & Ours       \\ \midrule
EOG                                   & 83.3 (2.2) & 83.0 (9.9)  & 89.2 (0.0)  & \textbf{93.0 (0.7)} & 68.1 (4.5)  & 57.5 (4.1)  & 83.9 (3.6)  & \textbf{91.9 (1.5)} \\
EEG                                   & 84.4 (2.1) & 75.6 (9.4)  & 89.2 (0.5)  & \textbf{92.6 (0.6)} & 68.9 (3.5)  & 51.5 (7.3)  & 84.6 (3.2)  & \textbf{91.3 (1.7)} \\
Resp                                  & 82.9 (1.9) & 87.3 (2.3)  & 86.3 (0.8)  & \textbf{91.4 (0.5)} & 73.1 (11.1) & 56.6 (10.4) & 86.3 (2.7)  & \textbf{88.3 (1.7) }\\
SpO\textsubscript{2} & 64.8 (1.4) & 63.2 (3.0)  & 83.5 (1.0)  & \textbf{91.5 (0.3) }& 51.0 (1.6)  & 75.1 (5.8)  & 83.5 (1.0)  & \textbf{89.8 (0.7)} \\
CO\textsubscript{2}  & 81.1 (3.4) & 80.4 (5.9)  & 89.4 (0.5)  & \textbf{92.9 (0.5) }& 73.3 (18.6) & 88.6 (0.7)  & 89.4 (0.5)  & \textbf{92.5 (0.9)} \\
ECG                                   & 80.8 (2.3) & 85.6 (2.3)  & 88.5 (0.5)  & \textbf{93.1 (0.6)} & 61.8 (6.9)  & 54.0 (8.8)  & 81.0 (3.6)  & \textbf{91.8 (0.8)} \\ \midrule
EOG, EEG                              & 79.5 (2.9) & 74.7 (12.2) & 88.7 (0.5)  & \textbf{91.5 (0.8)} & 62.3 (2.2)  & 47.8 (5.1)  & 82.7 (4.0)  & \textbf{90.3 (2.1)} \\ \midrule
None                                  & 86.4(1.2)  & 89.4(0.6)   & 90.4 (0.4)  & \textbf{93.4 (0.4)} & 86.8(1.0)   & 89.7(0.7)   & 90.0 (0.8)  & \textbf{93.3 (0.6)} \\ \bottomrule
\end{tabular}
}
\end{table*}

\section{Discussion}
\label{sec:disc}


In this study, we presented a machine learning pipeline for apnea detection with any combination of available PSG modalities. By leveraging the complementary nature of multiple modalities, data fusion techniques can mitigate the impact of missing data, reduce uncertainty, improve the overall robustness of the fused information, and compensate for the absence of certain modalities. All of these can allow for higher performance in machine learning tasks \citep{woo2023towards}.

We showed that our hybrid pipeline ––which combines a transformer-based architecture and a gated fusion–– is robust to noisy and missing modalities and outperforms several recent baselines. Specifically, we investigated the effect of imperfect (noisy and missing) signals on the performance of apnea detection models. The performance of the models that are not designed to handle such conditions can degrade drastically as the imperfection of input signals increases. As shown in Figures \ref{fig:missing} and \ref{fig:noisy}, our model is more resilient to various degrees of missingness and noise in comparison with the baselines.

In this study, we used six different modalities, each of which with various noise types. Some of these types of noise have been comprehensively studied, and efficient removal approaches are available for such noises \citep{limaye2016ecg,lai2018artifacts}. As a result, some standard noises are removable using well-known denoising approaches without losing too much information and performance degradation. For instance, the ``baseline wander" and ``power-line interference" noises in ECG signals are removable using a notch filter (a filter that eliminates a single frequency from a spectrum of frequencies) and a high-pass filter (a filter that passes signals with a frequency higher than a certain cutoff frequency). To demonstrate the robustness of our proposed pipeline to this type of noise, we report the performance of our method in the presence of such noise in Appendix \ref{apx:result}.

In our study, however, we focus on noises that can deeply disturb the signals so that the original signals are not distinguishable. This type of noise is not easily reproducible, like the easy-to-model noises we discussed above. Modeling this type of noise is very hard. Such noise types include patient movements or electrode detachment, which are events that can introduce severe noise.

The clinical implications of our study are also worth noting. Our method can be especially relevant to the ongoing efforts to adopt at-home sleep apnea testing (HSAT) solutions. While there exist several FDA-approved commercial HSAT products \citep{van2022multicentric,manoni2020new},  inside-clinic PSG is still the gold standard for diagnosing sleep apnea. This is partly due to the challenges that sleep recording can present. 

Research on HSATs is especially relevant to pediatric populations, as such tools are almost widely available for adults, but not children \citep{kirk2017american}. In fact, a key barrier to running sleep apnea testing outside the clinic for children is the presence of noise and missingness, as children can move more frequently, be less cooperative, and pull the probes, among other issues. 

Additionally, a key difference between apnea detection in adults versus children relates to using SpO\textsubscript{2} (for adults) versus EEG (for children) as the main signal for apnea detection \citep{bandla2000dynamic}. In children, the brain activates early on to regulate breathing and sleep disorders, therefore it functions as a better signal for apnea detection. As we show in Table \ref{tab:modality-missing}, even without EEG and EOG (i.e., the two critical but challenging to collect signals in children), our method performs very well. 

The two datasets that we use in this study are related to pediatric sleep studies, further showcasing the performance of our pipeline on pediatric cases. We also report the performance of our model across different ages in Appendix \ref{sec:age} to further highlight the potential of our pipeline in informing the efforts targeting younger age patients.

Some limitations of our study are worth noting. 
A common concern in apnea detection models relates to the lack of sub-typing. Although apnea and hypopnea have three different types (i.e., obstructive, central, and mixed) \citep{javaheri2017sleep}, our work, similar to most prior studies, only tries to train a model that detects apnea, not its type and severity. Additionally, while there are many explanation methods for opaque-box methods \citep{belle2021principles}, our method (an opaque-box one) would still offer limited explainability. 

Moreover, in this study, we manually added noise to the lab-recorded PSG data. While we are not aware of any large study collecting at-home sleep data, evaluating our model using data with actual noise could have demonstrated our model's performance more comprehensively.  

In the future, we plan to investigate whether applying our method can also help to improve sleep staging performance in comparison with existing methods. Sleep staging is mostly done using EEG and EOG. 
 \acks{Our study was supported by NIH awards, P20GM103446 and U54-GM104941.}

\bibliography{fayyaz23}

\newpage

\appendix

\section{Transformer Backbone} \label{tfb} 
The encoder and decoder consist of multiple layers of transformer, each consisting of two components. Both components have a residual connection and a normalization layer. The first is the multi-head self-attention (MHSA), which allows the different parts of the input sequence to interact:

\begin{equation}
 X^{'} = LayerNorm(X)
\end{equation}

\begin{equation}
MHSA(X^{'}) = Concat(h_{1}, ..., h_{n}) W^{C}
\end{equation}

\noindent
where:

\begin{equation}
 h_{i} = Softmax(\frac{W^{K}_{i}X^{'}(W^{N}_{i}X^{'})^{T}}{\sqrt{d_{k}}})W^{V}_iX^{'}.
\end{equation}

The second component is a fully connected network (FCN) which applies a nonlinear transformation to each position in the sequence:

\begin{equation}
    X^{''} = LayerNorm(X^{'}+MHSA(X^{'}))
\end{equation}

\begin{equation}
    FCN(X^{''}) = ReLU(X^{''}W_{1} + b_{1})W_{2} + b_{2},
\end{equation}

\begin{equation}
    output = X^{''} + FCN(X^{''})
\end{equation}

$W_{i}^{N}, W_{i}^{K}, W_{i}^{V}$ are learnable weights related to the head $i$ of the self-attention module. $d_{k}$ is the dimension of queries and keys in self-attention. ($W_{1}$, $b_{1}$) and ($W_{2}$, $b_{2}$) are the learnable weights and biases for the first and second layers, respectively.

\section{Additional training details}
\label{apd:train}

In experiments, we used Adam optimizer \citep{kingma2014adam} with a learning rate of $10^{-3}$, $\beta_{1}=0.9$, ${\beta}_{2}=0.999$, and $\epsilon=10^{-7}$. A batch size of 256 was used for training. We use L2 weight regularization with $\lambda=10^{-3}$ and dropout (rate=0.25) to avoid overfitting. We trained the model for 100 epochs in a 5-fold cross-validation manner. our model was implemented using  Keras \citep{chollet2015keras} inside the  TensorFlow \citep{abadi2016tensorflow} framework.

\paragraph{Hyper-parameters tuning} We also tested our model with different hyper-parameters. We ran the experiments with models with different numbers of transformer layers and heads in each layer. For our task, the best performance achieved with the model has 5 layers and 4 heads in each layer. in our architecture, each layer consists of two layers of MLP with 16 and 32 units. Fully connected layers in a gated fusion network have 8 units, which have been selected among 4, 8, 16, and 32 units.

\section{Extended Results}
\label{apx:result}

The performance of SOTA models in terms of AUROC with the concurrent occurrence of noise and missingness on the NCH dataset is shown in Table \ref{tab:noise-missing-sota}. The performance of the proposed model in terms of AUROC when just utilizing ECG signals with Baseline wandering and Powerline interference noises are shown in Table \ref{tab:apd-noise}.

\begin{table}[h]
    \centering
\begin{tabular}{l|cc}
\hline
\multirow{2}{*}{Noise type} & \multicolumn{2}{c}{AUROC} \\ \cline{2-3} 
                            & CHAT        & NCH         \\ \hline
Baseline wandering          & 82.2 (0.7)  & 82.5 (0.7)  \\
Powerline interference      & 82.4 (0.8)  & 82.8 (0.7)  \\
Without Noise               & 83.1 (0.6)  & 83.4 (0.6)  \\ \hline
\end{tabular}
    \caption{Performance of our model with noisy ECG signals. Mean($\pm$ STD)}
    \label{tab:apd-noise}
\end{table}

\begin{table*}[ht]
\centering
\caption{Performance of the baselines and our model in terms of AUROC with concurrent occurrence of noise and missingness on NCH dataset. The mean (standard deviation) values are shown.}
\label{tab:noise-missing-sota}
\begin{tabular}{cccccc}
\toprule
Missing ratio         & SNR & CNN-LSTM  & CNN       & Transformer & Ours      \\
\midrule
\multirow{5}{*}{10\%} & 10  & 57.0(3.3) & 58.8(2.0) & 82.1(0.2)   & 84.9(1.9) \\
                      & 20  & 66.6(5.5) & 73.7(1.1) & 83.0(0.3)   & 90.4(0.9) \\
                      & 30  & 80.4(0.9) & 80.5(0.7) & 84.6(0.5)   & 92.6(0.5) \\
                      & 40  & 81.8(0.9) & 81.4(1.0) & 86.3(0.5)   & 92.8(0.5) \\
                      & 50  & 81.8(1.0) & 81.2(0.9) & 86.5(0.3)   & 92.8(0.5) \\ \midrule
\multirow{5}{*}{20\%} & 10  & 57.4(2.4) & 58.8(1.7) & 79.4(0.2)   & 82.9(1.9) \\
                      & 20  & 64.6(3.7) & 69.0(0.7) & 80.0(0.5)   & 89.0(1.1) \\
                      & 30  & 74.9(0.5) & 74.1(0.5) & 81.8(0.4)   & 91.8(0.6) \\
                      & 40  & 75.9(0.7) & 74.7(0.9) & 83.1(0.4)   & 92.0(0.5) \\
                      & 50  & 75.9(0.9) & 74.5(1.0) & 83.3(0.4)   & 92.1(0.5) \\ \midrule
\multirow{5}{*}{30\%} & 10  & 56.8(2.0) & 58.6(1.2) & 76.2(0.5)   & 80.9(2.0) \\
                      & 20  & 62.3(2.2) & 65.2(0.6) & 77.1(0.3)   & 87.3(1.0) \\
                      & 30  & 70.2(0.6) & 68.7(0.5) & 78.3(0.5)   & 90.6(0.6) \\
                      & 40  & 70.5(1.2) & 69.3(0.8) & 79.5(0.5)   & 91.1(0.5) \\
                      & 50  & 70.4(1.2) & 69.3(0.8) & 80.0(0.5)   & 91.1(0.5) \\ \midrule
\multirow{5}{*}{40\%} & 10  & 56.6(0.9) & 57.6(0.9) & 73.1(0.7)   & 78.6(1.9) \\
                      & 20  & 61.0(1.6) & 62.4(0.5) & 73.8(0.9)   & 85.4(1.2) \\
                      & 30  & 66.0(0.4) & 64.7(0.3) & 74.8(0.8)   & 89.3(0.6) \\
                      & 40  & 66.3(1.1) & 65.0(0.7) & 76.1(1.0)   & 89.9(0.5) \\
                      & 50  & 66.4(1.1) & 64.9(0.9) & 76.3(1.1)   & 89.8(0.5) \\ \midrule
\multirow{5}{*}{50\%} & 10  & 56.0(1.0) & 56.6(0.9) & 69.7(0.9)   & 76.9(2.0) \\
                      & 20  & 58.6(1.4) & 59.7(0.5) & 70.4(0.9)   & 83.4(1.0) \\
                      & 30  & 62.3(0.9) & 60.9(0.4) & 71.3(0.8)   & 87.6(0.6) \\
                      & 40  & 62.6(0.8) & 61.1(0.7) & 72.3(1.2)   & 88.0(0.6) \\
                      & 50  & 62.5(0.4) & 61.2(0.7) & 72.7(1.2)   & 88.2(0.4) \\ 
\bottomrule
\end{tabular}
\end{table*}


\section{Noise and missingness algorithms} \label{algs}
 Additive white Gaussian noise generation and Random epoch-channel omission are shown in algorithms \ref{alg:noise} and \ref{alg:missing}, respectively.

\begin{algorithm2e}
\caption{Random epoch-channel omission}\label{alg:missing}
\KwIn{$\{S_{i}\}_{i=1}^{I}, omission\textunderscore ratio$}
\KwOut{$\{S_{i}\}_{i=1}^{I}$}
\For{$i \gets 1$ \textbf{to} $I$}{
    \For{$j \gets 1$ \textbf{to} $J_i$}{
        \For{$m \gets 1$ \textbf{to} $M$}{
             $rnd \leftarrow$ Generate a random number in (0,1) \\     
              \uIf{$rnd \leq omission\textunderscore ratio$}{
                    $X^{i,j}_{m} = 0$
                }
            }
        }
    }
\end{algorithm2e}

\begin{algorithm2e*}
\caption{Additive white Gaussian noise generation}\label{alg:noise}
\KwIn{$\{S_{i}\}_{i=1}^{I}, target\textunderscore snr, noise\textunderscore occurrence\textunderscore chance$}
\KwOut{$\{S_{i}\}_{i=1}^{I}$}
\For{$i \gets 1$ \textbf{to} $I$}{
    \For{$j \gets 1$ \textbf{to} $J_i$}{
        \For{$m \gets 1$ \textbf{to} $M$}{
                    $signal\textunderscore average\textunderscore power = \frac{\sum_{t=1}^{T}(X^{i,j}_{t,m})^2}{T}$ \\
                    $signal\textunderscore average\textunderscore power\textunderscore db = 10 * log_{10}(signal\textunderscore average\textunderscore power)$ \\
                    $noise\textunderscore average\textunderscore power\textunderscore db = signal\textunderscore average\textunderscore power\textunderscore db  - target\textunderscore snr$\\
                    $noise\textunderscore average\textunderscore power = 10^{noise\textunderscore average\textunderscore power\textunderscore db}$\\
                    $noise \sim Normal(0, \sqrt{noise\textunderscore average\textunderscore power})$\\
                     $rnd \leftarrow$ Generate a random number in (0,1) \\
                    \uIf{$rnd \leq noise\textunderscore occurrence\textunderscore chance$}{
                    $X^{i,j}_{m} \leftarrow X^{i,j}_{m} +  noise$\\
                }

            }
        }
    }
\end{algorithm2e*}

\section{Model performance across different ages}
\label{sec:age}

\begin{figure}[h]
    \centering
  \includegraphics[width=1\linewidth]{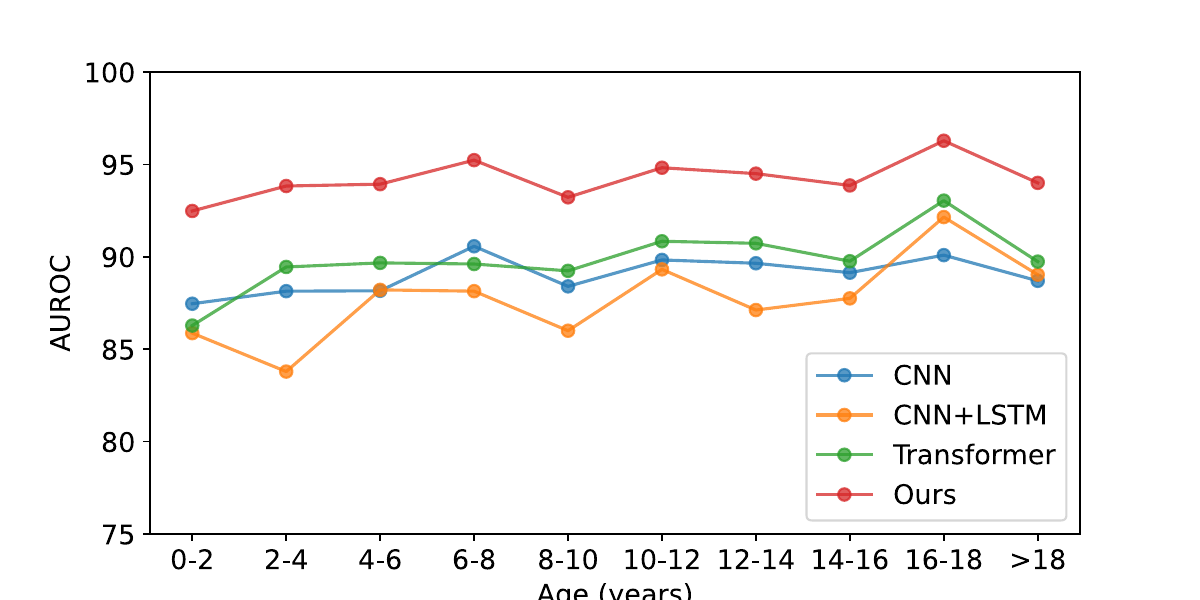}
    \caption{Performance of the proposed and state-of-the-art models across different age groups on the NCH dataset.}
    \label{fig:age}
\end{figure}

We separated out patients according to their age range to study the performance of the methods across different ages. The results for the NCH dataset are shown in Figure \ref{fig:age}. One can observe that the model's discriminative performance remains consistently over 90\%, while the performance is slightly lower in younger ages. The patients' age distribution is shown in Figure \ref{fig:age}.

\end{document}